# Behavior of Hydroxyl Radicals on Water Ice at Low Temperatures

Masashi Tsuge and Naoki Watanabe*

Institute of Low Temperature Science, Hokkaido University, N19-W8, Kita-ku, Sapporo, Hokkaido 060-0819, Japan

**Abstract**
**Conspectus**

Because chemical reactions on/in cosmic ice dust grains covered by amorphous solid water (ASW) play important roles in generating a variety of molecules, many experimental and theoretical studies have focused on the chemical processes occurring on the ASW surface. In laboratory experiments, conventional spectroscopic and mass-spectrometric detection of stable products is generally employed to deduce reaction channels and mechanisms. However, despite their importance, the details of chemical reactions involving reactive species (i.e., free radicals) have not been clarified because of the absence of experimental methods for *in situ* detection of radicals. Because OH radicals can be easily produced in interstellar conditions by not only the photolysis and/or ion bombardments of $H_2O$ but also the reaction of H and O atoms, they are thought to be one of the most abundant radicals on ice dust. In this context, the development of a close monitoring method of OH radicals on the ASW surface may help to elucidate the chemical reactions occurring on the ASW surface.

Recently, to detect OH radicals adsorbed on the ASW surface, we applied our developed method to sensitively and selectively detect surface adsorbates with a combination of photostimulated desorption and resonance-enhanced multiphoton ionization techniques. Using this method, we showed that an OH radical on the ASW surface can be desorbed upon one-photon absorption at 532 nm, at which wavelength both the OH radical and $H_2O$ molecule are transparent. Theoretical calculations addressing an OH radical adsorbed on water clusters indicated that the valence A–X transition of an OH radical significantly red-shifts by ~2 eV when the OH radical is strongly adsorbed to ice through three hydrogen bonds. With this method, the number density of surface OH can be monitored as a snapshot so that the behaviors of OH radicals, such as surface diffusion, can be studied. Moreover, the development of a system for studying the wavelength dependence of photodesorption may establish a foundation for future research investigating the absorption spectra of surface adsorbed species.

Owing to the large electron affinity of OH radicals on ice, they are expected to easily become OH¯ by electron attachment on the ASW surface. We characterized the behavior of OH¯ on ASW at low temperatures, which may be relevant not only to physicochemical processes on cosmic ice dust and planetary atmosphere but also to understanding the electrochemical properties of ice. A negative constant current was found when ASW at temperatures below 50 K was exposed to both UV photons and electrons. It was demonstrated that the negative current is initiated by the formation of OH¯ ions on the ASW surface, and they are transported to the bulk via the proton-hole transfer mechanism, which was predicted 100 years ago as a mirror image of proton transfer known as the Grotthuss mechanism. These results indicate that OH¯ ions are readily transported to the bulk ice and further induce reactions, even at low temperatures where thermal diffusion is negligible. Therefore, in-mantle chemical processes that have been considered inactive at low temperatures are worth reevaluating.

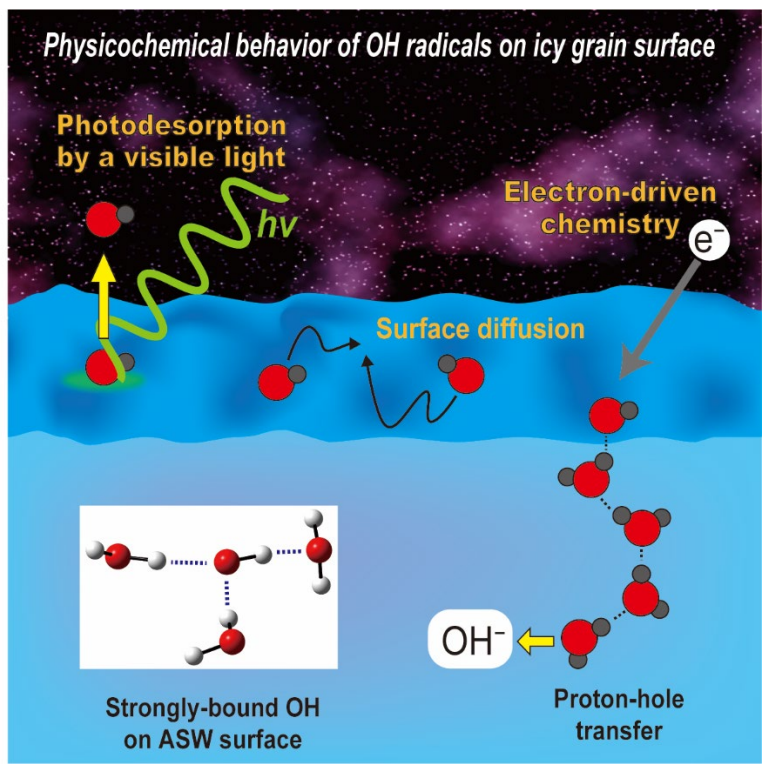

## ■ KEY REFERENCES

- Miyazaki, A.; Watanabe, N.; Sameera, W. M. C.; Nakai, Y.; Tsuge, M.; Hama, T.; Hidaka, H.; Kouchi, A. Photostimulated desorption of OH radicals from amorphous solid water: How visible light interacts with an OH-ice complex. *Phys. Rev. A* **2020**, *102*, 052822.[1] *Experimental and theoretical studies suggested that the photodesorption of OH radicals from the ASW surface by a visible photon is initiated by the one-photon absorption of strongly adsorbed OH. The methods and possible mechanisms are described in detail.*

- Watanabe, N.; Sameera, W. M. C.; Hidaka, H.; Miyazaki, A.; Kouchi, A. Ultraviolet-photon exposure stimulates negative current conductivity in amorphous ice below 50 K. *Chem. Phys. Lett.* **2019**, *737*, 136820.[2] *The negative current conductivity of ASW below 50 K was identified for the first time. Experimental and theoretical approaches indicated that the current was induced by proton-hole transfer, the occurrence of which has been an open question for 100 years.*

- Kitajima, K.; Nakai, Y.; Sameera, W. M. C.; Tsuge, M.; Miyazaki, A.; Hidaka, H.; Kouchi, A.; Watanabe, N. Delivery of Electrons by Proton-Hole Transfer in Ice at 10 K: Role of Surface OH Radicals. *J. Phys. Chem. Lett*, **2021**, *12*, 704.[3] *Combining the current measurements and PSD-REMPI detection of surface OH radicals, the mechanism of negative current conductivity through ASW was revealed to be proton-hole transfer initiated by the capture of electrons by OH radicals.*

## 1. Introduction

In very cold and dense interstellar regions, so-called molecular clouds (MCs), a large variety of molecules, up to 150 species, has been identified,[4] meaning that chemical reactions occur even at temperatures as low as 10 K. The history of the molecular formation towards large molecules is known as chemical evolution in MCs. Ion-molecule and radical-radical reactions in the gaseous phase can efficiently occur, even at low temperatures, but the large complexity and abundance of those molecules cannot be explained by only considering gas phase chemistry.[5,6] Therefore, it has been considered that chemical processes on a low-temperature dust surface play important roles in chemical evolution. In MCs, the surface of cosmic dust is typically covered with so-called ice mantle, which is predominantly composed of water ice in the amorphous phase (amorphous solid water, ASW) with other molecules, such as CO, $CO_2$, $NH_3$, $CH_4$, $H_2CO$, and $CH_3OH$.[7] Because of the relevance of ASW, many laboratory and theoretical studies have focused on chemical processes occurring on the ASW surface. An important example of experimental findings is confirming a formation route for $H_2CO$ and $CH_3OH$ under the condition of MCs. Several groups demonstrated that the successive hydrogenation of CO on the ASW surface,

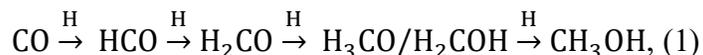

$$CO \xrightarrow{H} HCO \xrightarrow{H} H_2CO \xrightarrow{H} H_3CO/H_2COH \xrightarrow{H} CH_3OH, (1)$$

efficiently proceeds, even at a low temperature of ~10 K.[5,8-10] Because of reaction barriers in the first and third steps, thermally activated reactions do not occur at low temperatures. However, this sequential reaction was found to proceed even at low temperatures via quantum tunneling within an experimental timescale. The successive hydrogenation process is adopted as a plausible pathway to form various interstellar organic molecules, especially in the dense core of MCs where radiation fields are weak.[11,12] In the low-temperature dense regions of MCs, chemical evolution in the ice mantle is dominated by the surface hydrogenation reactions of atoms and simple molecules accreted from the gas phase. However, note that hydrogenation alone cannot produce complex organic molecules (COMs), which require reactions between heavier species, including radicals. In contrast to hydrogen atoms, heavier species cannot move at approximately 10 K and should accumulate on the ice surface. These accumulated heavier radicals are key for the production of COMs.

During star formation, the dust and gas temperature gradually increase from 10 to 200 K to produce a hot-core region. As the temperature of dust rises, heavier species start to diffuse, encounter the other adsorbates on ice dust, and eventually evaporate into the gas phase. Diffusive reactions of radicals such as OH, HCO, $CH_3O$, $CH_2OH$, NH, and $NH_2$ at this stage lead to a wide variety of COMs[13] through barrier-less radical reactions, as listed in Figures 1 and 2 of ref 13. Among interstellar radicals, hydroxyl radicals (OH) are thought to be very abundant as photoproducts and precursors of $H_2O$ on the ice mantle. Accordingly, OH radical reactions on the ASW surface, $OH + CO \rightarrow CO_2 + H$,[14] $OH + H_2 \rightarrow H_2O + H$,[15] and $OH + OH \rightarrow H_2O_2$ (and $H_2O + H$),[16] have been studied. These studies were performed by spectroscopically measuring stable molecular products on ice. Another experimental technique often used in laboratory investigations is the temperature-programmed desorption method, in which adsorbates sublimating during warming up are detected by mass spectrometry. In these experiments, information about surface diffusion prior to reactions cannot be provided. To evaluate the role of OH radicals in chemical evolution, it is crucial to understand how OH radicals behave on the ice surface. To obtain further insight into the physicochemical behavior of OH radicals on ice, *in situ* detection is particularly important. However, conventional methods, such as infrared, Raman, and electron spin resonance spectroscopies, are not always appropriate for this purpose because spectroscopic discrimination between surface and bulk species is sometimes difficult. Furthermore, an intrinsic difficulty arises from the high reactivity of radicals. That is, it is difficult to accumulate enough reactive radicals on a surface for detection. Although surface sensitive detection is in principle possible using microscopic techniques such as scanning tunneling microscopy and field-emission microscopy, these methods are not appropriate for distinguishing OH radicals from $H_2O$.

Recently, our group developed a new method to sensitively detect surface adsorbates with a combination of photostimulated desorption (PSD) and resonance-enhanced multiphoton ionization (REMPI) techniques. This PSD-REMPI method has been successfully applied for detecting atomic and molecular hydrogen on analogues of astronomical dust surfaces,[17-22] and the experimental outcomes are summarized in a recent review paper.[23] In this account, we introduce the application of the PSD-REMPI method for the detection of OH radicals on ASW surfaces (Section 2).[1] In Section 3, the electron-driven chemistry of OH radicals on ASW, as demonstrated from the observation of a negative charge current, will be overviewed,[2,3] and a future outlook will be given in Section 4.

## 2. OH radicals on ASW: *in situ* detection
### 2.1 PSD-REMPI method

In this section, we will introduce a method for *in situ* surface adsorbate detection using the PSD-REMPI method. In this method, the amount of photodesorbed surface adsorbates is measured as an ion signal, which is proportional to the surface number density. The REMPI technique and time-of-flight mass spectrometry (TOF-MS) enable sensitive, species-selective, and state-selective detection. Since the PSD-REMPI signal is acquired as a snapshot, the time evolution of surface number densities for adsorbates can be studied with this method. Figure 1 shows a typical setup and timing chart for the PSD-REMPI measurements. Typically, an aluminum substrate connected to the cold head of the closed-cycle helium refrigerator is located at the center of the ultrahigh vacuum chamber. For the OH radical experiments, ASW was produced on the substrate at 100 K by backfilling $H_2O$ vapor, and ASW was thus continuously exposed to photons from a deuterium lamp. The surface adsorbates were photodesorbed by unfocused weak nanosecond laser irradiation (PSD laser), typically <100 μJ per pulse at 532 nm. After a certain delay of the PSD laser shot, the desorbed species were selectively ionized by the REMPI laser, which was typically focused 1 mm above the substrate, and the produced ions were detected using TOF-MS. From the delay and distance between the focal point of the REMPI laser and the surface, one can also determine the translational energy distribution of photodesorbed species.

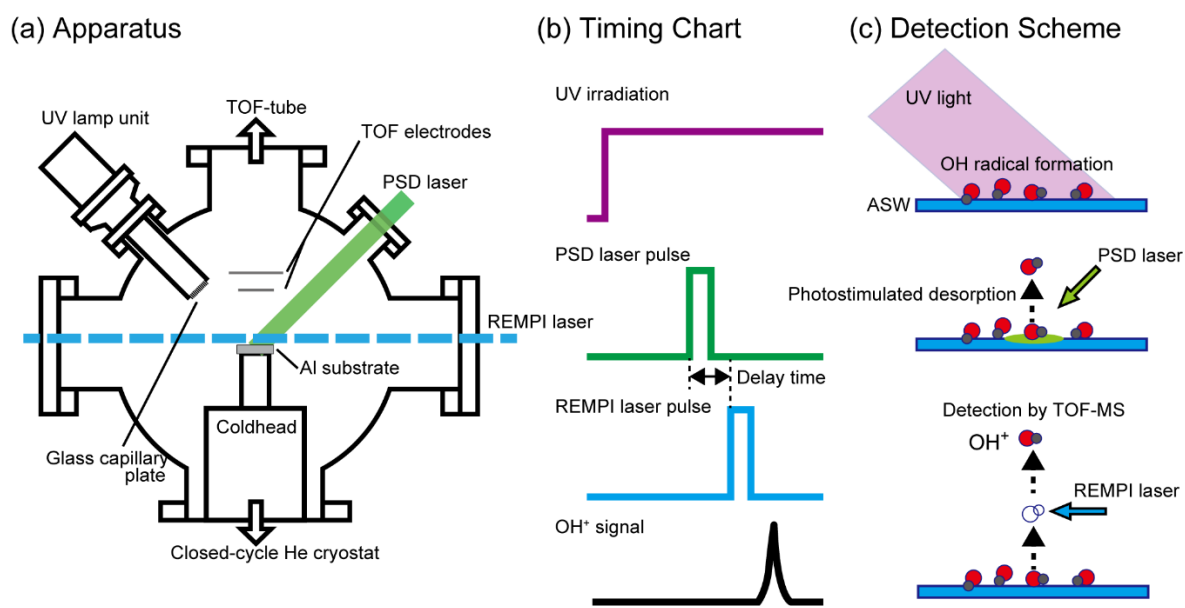

**Figure 1.** (a) Experimental setup, (b) timing chart, and (c) detection scheme for the PSD-REMPI measurement of OH radicals.

## 2.2 Application for OH radicals on ASW[1]

Hydroxyl radicals were produced by irradiating the ASW sample by the deuterium UV lamp. The UV photons mainly induce the dissociation of $H_2O$ into H + OH, and at the experimental temperature (70 K) none of the possible photoproducts (H, $H_2$, O, and $O_2$) other than OH radicals can stay on the surface. During UV irradiation, the produced OH radicals were desorbed by a PSD laser and subsequently ionized by the REMPI laser. The REMPI laser wavelength was tuned such that the OH radicals are ionized by the (2+1) REMPI scheme via the intermediate D $^2\Sigma^-$ state;[24] a rotationally-resolved REMPI spectrum is shown in Figure 2. Hydrogen peroxide ($H_2O_2$) is also produced upon UV irradiation of water ice. However, the detected OH radicals are not originated from the photodissociation of $H_2O_2$ by PSD laser irradiation because the absorption cross section of $H_2O_2$ at 532 nm is negligibly small.

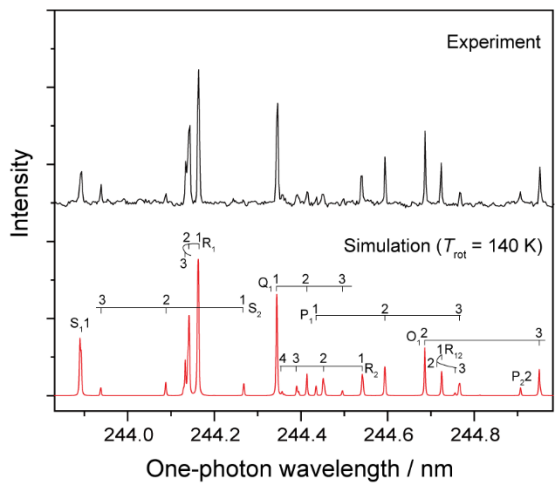

**Figure 2.** PSD-REMPI spectrum of OH desorbed from an ASW surface (top) and the simulated spectrum of the transition D $^2\Sigma^-$ ($v' = 0$) ← X $^2\Pi$ ($v'' = 0$) with a rotational temperature of 140 K (bottom).

Figure 3(a) shows the variation in the OH signal intensities at a 244.164 nm REMPI-laser wavelength for the R1(1) branch of a transition, D $^2\Sigma^-$ ($v' = 0$) ← X $^2\Pi$ ($v'' = 0$), as a function of the delay time between PSD and REMPI laser pulses. The signal intensity distribution shows two peaks, with the faster component near 1 μs delay time and the slower component near 3 μs. The observed profile is directly correlated with the translational energy distribution of photodesorbed OH radicals.[25] Fitting the profile with the Maxwell-Boltzmann distribution gives two components with translational temperatures of 3000 ± 600 and 320 ± 70 K, which correspond to 0.26 ± 0.05 and 0.027 ± 0.006 eV, respectively. These translational temperatures were independent of the ASW surface temperature in the range of 10–70 K. The profile of the Maxwell-Boltzmann distribution does not necessarily mean that desorption occurs at thermal equilibrium.[26] The observed rotational temperature of ~140 K (see Figure

2) was different from the translational temperatures, indicating a nonthermal photodesorption process. In particular, the faster component with a very high translational temperature of ~3000 K cannot be explained by laser-induced thermal desorption, while the slower component may be due to gentle desorption, such as phonon-mediated desorption.

To further clarify the photodesorption mechanisms, the PSD laser power dependence of each component was investigated, as shown in Figure 3(b). The intensity of the faster component showed a linear relationship with the PSD laser power, indicating that the faster component originates from a one-photon chemical process. On the other hand, the slower component showed a power-law dependence on the PSD laser power, where a similar power-law dependence was reported for the phonon-mediated desorption process.[27] In another experiment, the dependence of the intensity on ice thickness was investigated. The intensity of the faster component was only weakly dependent on ice thickness, consistent with photochemical desorption from the surface, while that of the slower component was significantly suppressed for thicker ice in accordance with the desorption mediated by the propagation of phonons created on the metal surface.

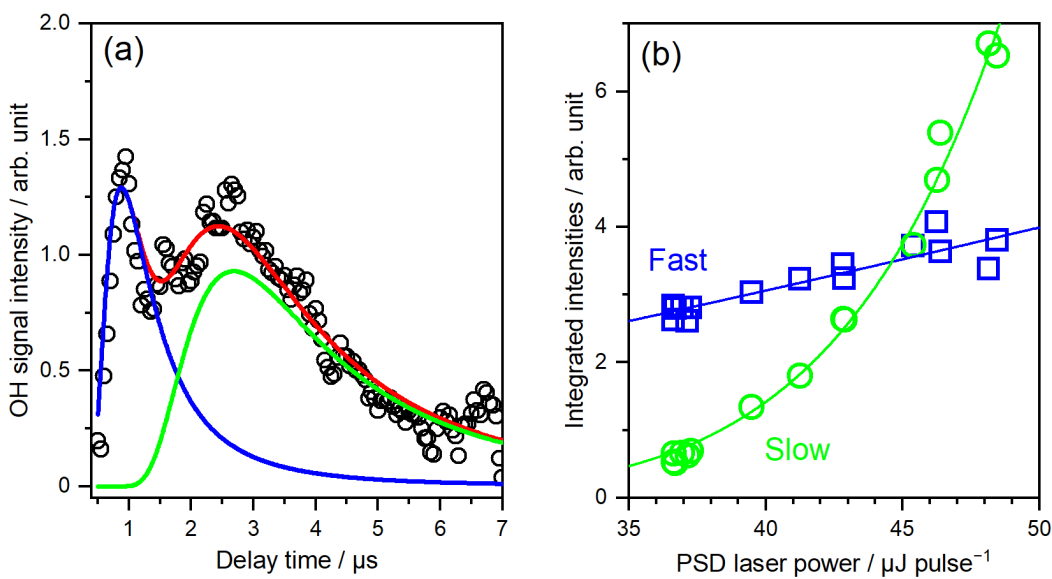

**Figure 3.** (a) Translational energy distribution of OH photodesorbed from an ASW surface at 70 K. Open circles represent the 5-point adjacent average of the obtained OH REMPI intensities. The red line is a fitting result with a two-component

Maxwell-Boltzmann distribution at 3000 K and 320 K for faster (blue) and slower (green) components. (b) The PSD laser power dependence of the intensities of faster and slower components.

Hereafter, we focus on the one-photon desorption mechanism of the faster component. Notably, neither the $H_2O$, OH, nor $H_2O$-OH complex absorbs near 532 nm, which is the wavelength of the PSD laser. As discussed in ref. 1, other possible desorption mechanisms, i.e., metal substrate-mediated processes such as electron-driven chemistry and plasmon-stimulated desorption, can be ruled out. Therefore, it is anticipated that OH radicals that adsorbed on the ASW surface absorb a photon at 532 nm, leading to desorption. We have performed quantum chemical calculations to obtain more insights into the OH radical behavior on ice. The two-layer ONIOM (our own N-layer integrated molecular orbital molecular mechanics) method[28] was employed to search the OH binding sites on ASW and calculate the vertical excitation energies of adsorbed OH radicals. The ASW model used in this work comprises 159 water molecules, where 46 water molecules are in the ONIOM high layer as the QM region, and the rest are in the low layer as the MM region (Figure 4(a)).

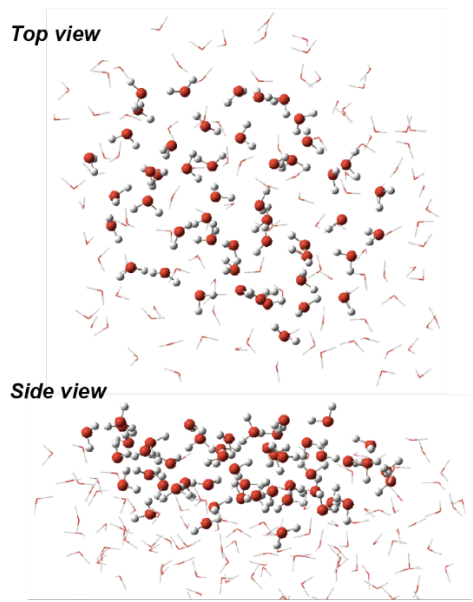

(a) Ice cluster model for ASW

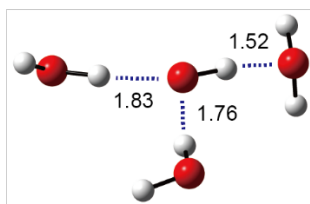

(b) Local structure of strongly bound OH

**Figure 4.** (a) Ice cluster model for ASW. Water molecules in the QM and MM regions are shown by "ball and stick" and "wireframe", respectively. (b) Local structure of the OH radical strongly bound to ASW. Dotted lines represent hydrogen bonds, and bond lengths are in Å.

By using this ASW model, eighteen kinds of binding sites for OH radicals were found with a range of binding energies of 0.06–0.61 eV. Binding energies are calculated as the difference between total energy of OH-adsorbed ice and the sum of energies of free OH and ice. Note that our calculations do not cover all the possible binding sites on ASW, because the structure of the ASW surface should change by ice preparation (both in the experiment and calculation). When OH radicals do not interact with dangling atoms on ice, the binding energy becomes as low as 0.06 eV. Such weakly bound OH radicals may be desorbed by a mild process, such as the propagation of phonons from the aluminum substrate. A very large binding energy, 0.61 eV, was found for the OH radical interacting with ice through three hydrogen bonds; the local structure of the strongly bound OH radical is shown in Figure 4(b). Hydroxyl radicals strongly bound on ice cannot be desorbed through a mild thermal process. To consider photochemical processes, we studied the excited states of the OH radicals on the ice surface.

The ground state of the isolated OH radical, $^2\Pi$, is doubly degenerated because there are two equivalent p-orbitals perpendicular to the molecular axis. When OH radicals interact with other species, this degeneracy will be lifted.[29] However, the energy splitting is so small (~0.02 eV) that the transition between these states should not be related to the photodesorption process induced by a 532 nm photon (2.33 eV). Thus, we focus on the valence transition $^2\Sigma^+ \leftarrow {}^2\Pi$, which corresponds to the A–X transition of isolated OH. Crawford et al. suggested, based on ab initio calculations, that the vertical A–X transition energy of the $H_2O$-OH complex (~3.8 eV) is red-shifted from that of isolated OH (~4.1 eV).[30] Computational studies on the A–X excitation energy for the $(H_2O)_n$-OH clusters have been limited to the size up to $n = 16$.[31,32] We found that a further red shift is induced upon increasing the number of water molecules surrounding OH and that the excitation energy converges when $n$ becomes approximately 20.

Figure 5 shows the A–X vertical excitation energy for the OH radical bound to the ASW surface calculated by TDDFT using the ωB97X-D/def2-TZVP level. In general, the calculated excitation energies fall in the range of 3–4 eV and are similar to those predicted for the $H_2O$-OH complex.[30] The A–X excitation of strongly bound OH radicals with a binding energy of 0.61 eV was found to occur at 580 nm (2.14 eV). Although this calculated A–X excitation energy does not exactly match the photon energy at 532 nm, OH radicals at some binding sites on the ASW surface certainly absorb the photon at 532 nm because the A–X excitation energy is very sensitive to the surface structure. In other words, the present PSD-REMPI method utilizing 532 nm PSD laser can detect, as a fast component, only OH radicals strongly bound to the ASW surface.

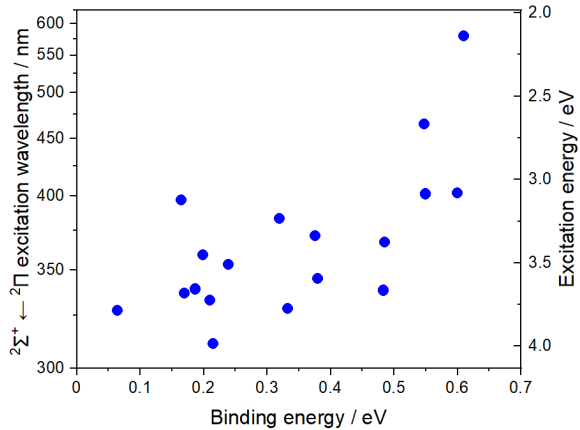

**Figure 5.** $^2\Sigma^+ \leftarrow {}^2\Pi$ vertical excitation wavelengths as a function of binding energies, calculated for OH radicals bound to the ASW surface.

Although qualitative determination of the desorption mechanism has yet to be done, we have drawn a possible mechanism based on preliminary calculations performed with a relatively small cluster containing 19 $H_2O$ molecules (see Figure 11 of ref. 1). One conical intersection was found near the minimum of the excited state (i.e., A state). Transition to another electronic state via conical intersection might lead to desorption of the photoexcited OH radical. Thus, the observed high translational energy can be due to the energy gap between the conical intersection and the dissociation limit. Simulating dissociation dynamics using the ice cluster model (presented in Figure 4(a)) is beyond the computational feasibility at this moment. Full-QM calculations performed on smaller $(H_2O)_n$-OH clusters will be feasible; the local structure of strongly bound OH radicals on ASW (see Figure 4(b)) can be constructed with a caged $(H_2O)_5$-OH cluster. TDDFT calculations predict that the A–X transition in $(H_2O)_5$-OH is also significantly red-shifted from the isolated OH. Therefore, the calculation of $(H_2O)_n$-OH clusters with a sufficiently high level of theory would provide further insights into photodesorption dynamics as well as the quantum chemical interpretation for a significant red shift.

## 3. Electron-driven chemistry of OH radicals on ice
### 3.1 Electrical conductivity of water ice

Diffusion of hydronium ($H_3O^+$) and hydroxide ($OH^-$) ions in aqueous solution is of fundamental importance because of its relevance to various chemical and biological processes.[33] These ions are known to rapidly diffuse via a relay of protons through a hydrogen bonded water wire: the so-called Grotthuss mechanism.[33,34] The positive current conductivity in water ice can also be described by the Grotthuss mechanism, although the detailed mechanism, especially at low temperature, is still under debate. A mirror image of the Grotthuss mechanism conducting a negative current is the concept of proton-hole transfer (PHT), which is the relay of the proton abstraction by a hydroxide ion from neighboring water: $OH^- + H_2O \rightarrow [OH^-\cdots H^+\cdots OH^-] \rightarrow H_2O + OH^-$. The PHT mechanism was first proposed approximately 100 years ago;[35] however, no clear evidence of the occurrence of this mechanism has been provided. For example, the mechanism of $OH^-$ migration in aqueous solution was found to be different from the PHT concept. In a few papers, the negative current conductivity of water ice was investigated.[36,37] In one study, $OH^-$ ions prepared in bulk by using base-doped ice were found to contribute to a small negative charge delivery at temperatures above 154 K; however, the mechanism was proposed to be the Brownian motion of $OH^-$, rather than the PHT mechanism.[36] How $OH^-$ ions doped on ice surfaces behave has not been investigated. Such behavior is relevant to the physicochemical processes on ice dust surfaces in space and planetary atmosphere, where the fate of the negative charge on OH, whether it is localized on the surface or transported to the bulk, would be an important issue.

In 2019, we found that ultraviolet-photon irradiation of ASW at temperatures below 50 K induces negative current conductivity and, based on quantum chemical calculations, is attributed to almost barrier-less PHT from the vacuum-ice interface to the metal substrate.[2] The details of the findings will be presented in Section 3.2. Recently, we have performed further experiments to clarify the mechanism of negative current delivery through ASW more explicitly. In Section 3.3, we will introduce some new findings, which certainly support the negative current conductivity results from the PHT mechanism.[3]

## 3.2 Identification of negative current conductivity in ASW[2]

An ASW sample with 40 or 120 monolayers (MLs) was produced over a nickel-plated sapphire disk at 10 K. The current through the ice was measured at the nickel surface using a picoammeter. In the first experiment depicted in Figure 6(a), the ASW sample kept at 10–50 K was irradiated with a deuterium lamp through a cylindrical UV light guide made of stainless steel such that UV photons illuminated the sample area only. Because UV photons create photoelectrons inside the metal guide, the ASW sample will be illuminated simultaneously with UV photons and these electrons. As shown in Figure 6(b), a negative constant current was observed during the operation of the UV lamp. This was the first finding on the negative current conductivity of ASW below 50 K. We note that transmission of electron is significantly suppressed at temperatures below 50 K.[37-39] To further elucidate the role of UV photons, the second experiment depicted in Figure 6(c) was performed. In this experiment, UV photons were supplied through a glass capillary plate mounted at the top of the guide to block photoelectrons generated within the guide, and electrons were separately provided by an electron gun placed in front of the substrate. The experimental result is shown in Figure 6(d). When only the electron gun was operated, the negative current immediately decreased. As previously reported,[37] it is possibly due to charging of the ASW sample. When UV photons were supplied under continuous exposure to electrons, a significant negative constant current was observed. This result strongly indicates that the negative constant current is induced by UV irradiation with the supply of electrons to the ASW surface. Because irradiation of ASW with the deuterium lamp primarily produces OH + H fragments, OH radicals are thought to play an important role in producing a negative current. Other possible photoproducts, such as H, $H_2$, O, and $O_2$, are not responsible for the observed negative current because the current was observed even at temperatures higher than desorption temperatures of these species. When an OH radical is adsorbed on the ASW surface, an electron that lands on the surface will preferentially combine with OH to produce $OH^-$ because the electron affinity of the OH radical (~1.8 eV) is larger than those of $H_2O$ and other photofragments. Therefore, these experimental results strongly indicated that the formation of $OH^-$ ions on the surface of ASW induces a negative current by the PHT mechanism.

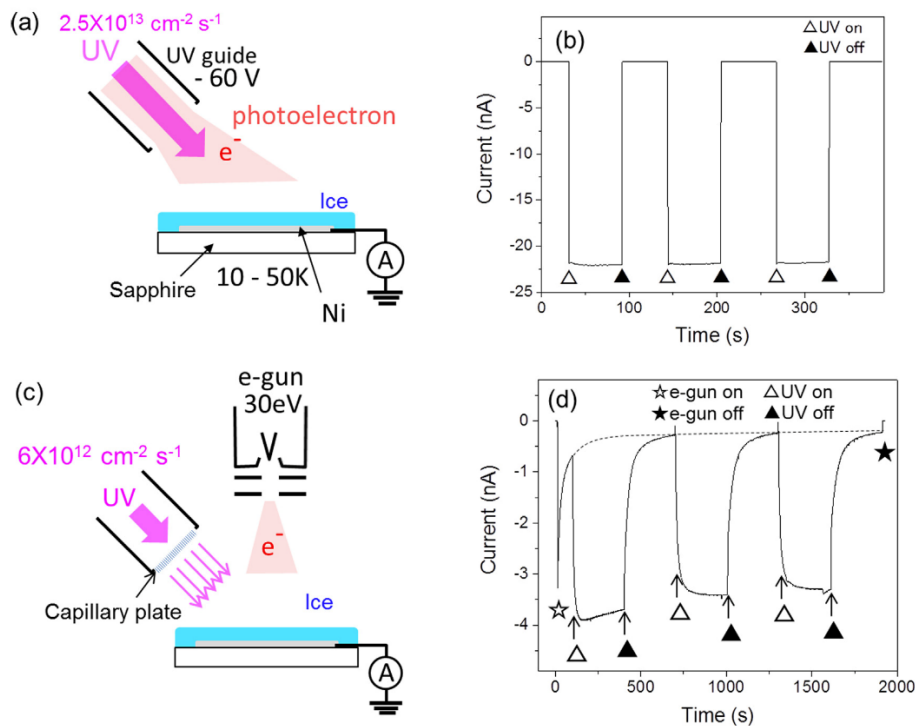

**Figure 6.** (a) Schematic of the first experiment and (b) temporal variation in the current measured through the 40 MLs of ASW at 20 K. (c) Schematic of the second experiment and (d) temporal variation in the current measured through the 40 MLs of ASW at 15 K. The dashed line is the current level only due to the electron gun without UV irradiation. Reproduced with permission from ref (2). Copyright 2019 Elsevier.

To examine the PHT mechanism in ice, quantum chemical calculations were performed using ice cluster models for ASW (similar to the one depicted in Figure 4(a)) and hexagonal water ice $I_h$. In both cases, the atom-centered density matrix propagation (ADMP) trajectory calculation[40] showed that the negative charge, which is initially localized as $OH^-$ on the surface of water ice, is rapidly transported to the bulk of ice through the PHT mechanism. The stationary points in the PHT reaction path were located using the ONIOM(QM:MM) calculation, and the result for the hexagonal water ice model is presented in Figure 7. These calculations indicate that $OH^-$ on the surface of water ice is energetically unfavorable compared with the structure where OH⁻ is in the bulk and that the potential minima are connected without significant barriers. These results are consistent with the absence of temperature dependence.[2] We presented the results of the hexagonal water ice model in Figure 7 for clarity, but qualitatively similar results were obtained for the ASW model (see Figure 4 of Ref. 2).

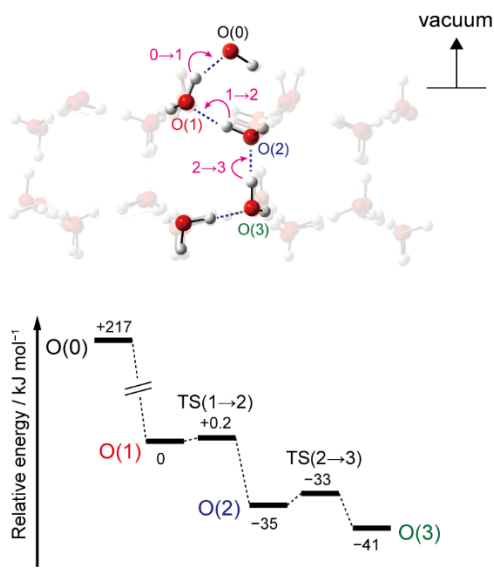

**Figure 7.** (Top) Slice of the hexagonal water ice cluster model with OH⁻ on the top. The movement of protons at each step of PHT is indicated by curved arrows. (Bottom) The potential energy diagram for three PHT steps. The labels O(0)–O(3) refer to the location of the OH⁻ radical in the hydrogen bonded chain depicted in the top layer. The total energy after the first PHT step (i.e., O(1)) is used as a reference, and the relative energies of the minima and transition states (TSs) are indicated by numbers.

### 3.3 Experimental identification of the carrier[3]

In the above experiments, if the OH photoproducts are consumed by OH⁻ production and subsequent PHT, the decrease in the amount of surface OH should be directly monitored by the PSD-REMPI method introduced in Section 2.1. The experiments were performed using an experimental setup similar to the one presented in Figure 1, with the addition of an electron gun. The ASW sample was deposited over a nickel-plated sapphire disk. The OH radicals were produced continuously by UV photons throughout the experiment. The wavelength of the REMPI laser and the time delay between PSD and REMPI lasers were adjusted so that the OH$^+$ signal intensity was maximized. A typical experimental result is shown in Figure 8. When the ASW sample was exposed to electrons, a negative current (panel a) and a depletion in the OH$^+$ signal ([OH$^+$]) (panel b) were observed, indicating that OH radicals at the surface are consumed upon exposure to electrons, leading to the negative current. Larger negative currents and OH radical depletions were observed for a larger incident current of electrons. When we define the degree of OH radical depletion as $(1 - [OH^+]_{e\text{-gun\_on}}/[OH^+]_{e\text{-gun\_off}})$, a clear correlation was seen between this value and the negative current density induced by UV irradiation (Figure 9). This result strongly indicates that OH radicals are involved in the generation of negative currents.

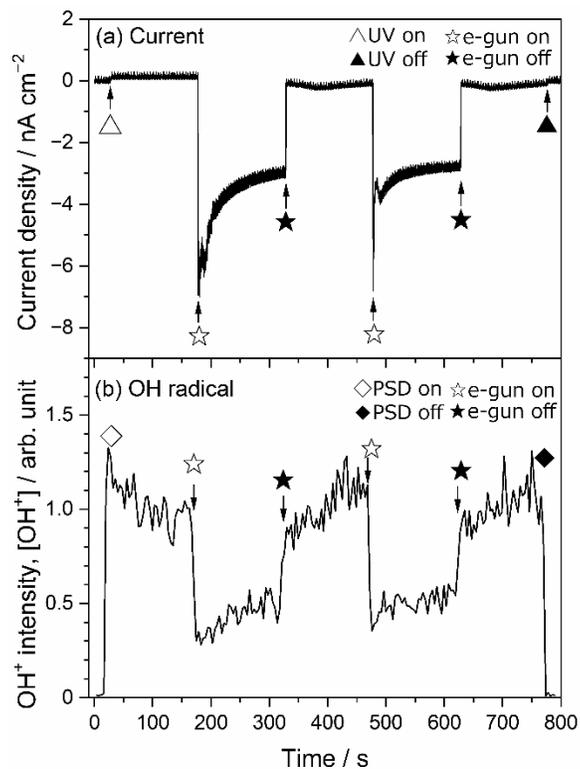

**Figure 8.** Behaviors of (a) the current through the ASW sample and (b) the surface number density of OH radicals upon exposure to UV photons and electrons. These data were obtained in separate measurements. During measurements of the surface number density of OH radicals, the ASW sample was continuously exposed to UV photons and, therefore, the OH$^+$ intensity recovers immediately after e-gun off.

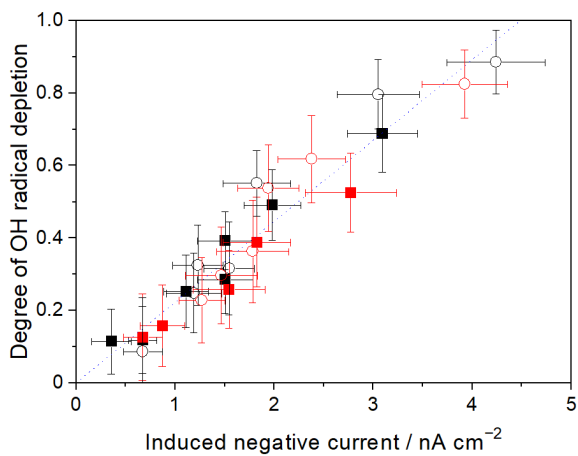

**Figure 9.** Degree of OH radical depletion (defined in text) as a function of the negative current induced by UV photons. The correlation was measured by varying the incident electron current. Open circles and filled squares represent results from two independent experiments, and black and red colors were used to identify rising and descending phases, respectively, of the incident current. The dotted line is a fit to all the data assuming a proportional relationship.

To further confirm that OH$^-$ ions are the main carrier of the negative current, we performed experiments using layered ice, where an ASW layer was grown over solid Ar on the nickel surface. When this ice is exposed to both UV photons and electrons, no enhancement of the negative current was observed. This result is reasonably explained by the fact that the negative current conductivity in ASW is induced by the PHT mechanism. OH$^-$ ions would be transported to the ASW/Ar interface, but because of the negative electron affinity of Ar,[41] electron transfer from OH$^-$ to Ar cannot occur. In other words, electron itself is not the carrier of observed constant negative currents and the release of trapped electrons by UV photons has a negligible contribution. In ref. 3, an ice-thickness dependence of the negative current as well as computational results on a thicker ice model (21 Å thick ice on the Ni(001) surface) are also presented.

## 4. Outlook
### 4.1 Photodesorption of free radicals adsorbed on ice

As introduced in Section 2.2, a combined experimental and theoretical approach indicates that OH radicals strongly bound to ASW can absorb a photon in a visible region, leading to desorption. It is interesting that when the OH radical strongly bonds to water ice, the photoexcitation energy of OH adsorbates red-shifts significantly and finally reaches energies in the visible region where isolated $H_2O$ and OH are transparent. According to the calculations, water molecules of up to 20 surrounding OH contribute to this drastic change in excitation energy. This kind of red shift may occur for other adsorbates, especially those forming hydrogen bonds with the ice surface. Further studies on both photoabsorption and desorption processes due to the red shift are of fundamental importance.

We are setting up a system for studying photodesorption at different wavelengths in the visible region. This work not only provides further insights into the photodesorption of OH radicals but also serves to develop a method for measuring the absorption spectrum of surface adsorbed species. High-sensitivity and species-selective detection by the PSD-REMPI method enables us to measure photodesorption spectra, even if several kinds of species are adsorbed on a surface.

Quantum chemical calculations are critically important for elucidating the detailed mechanism of the desorption process. For example, in the gas phase, the experimentally measured kinetic energy of the H-atom photoproduct from phenol has been well reproduced by quantum chemical calculations.[42] However, as mentioned earlier, simulating the dynamics of the desorption process for a large ice cluster (e.g., Figure 4(a)) with high accuracy is quite demanding, especially because the potential energy surface of electronically excited states should be treated. Thus, an alternative approach is to perform calculations for smaller $(H_2O)_n$-OH clusters with the full-QM method.

Because the PSD-REMPI method can *in situ* monitor the surface number density of OH radicals on ice, it can be applied to study the surface diffusion of OH radicals. Conventional techniques such as infrared spectroscopy and temperature-programmed desorption methods hardly approach the diffusion process. We first applied the PSD-REMPI method to determine the activation energy for OH diffusion on ASW.[43] In brief, the decay rates of PSD-REMPI signal intensities were measured at several temperatures, at which a decrease in the number density could be mostly due to recombination reactions, OH + OH → $H_2O_2$, and the activation energy was determined from an Arrhenius plot.

As mentioned in the Introduction, many kinds of free radicals play important roles in the formation of COMs on ice dust surfaces.[13] Relatively light radical species, such as OH, HCO, $CH_2$, $CH_3$, NH, and $NH_2$, start diffusing on the ice surface with increasing temperature. However, quantitative information on thermally activated diffusion and desorption has yet to be studied because of experimental difficulties. Therefore, the close monitoring of these species by the PSD-REMPI method is beneficial for drawing a detailed chemical model on ice surfaces. Studying the photodesorption processes of these radical species themselves would be important for modeling the gaseous phase chemical evolution. Unfortunately, the photodesorption processes of these radicals are still unknown. Interaction with visible light has not been carefully considered in the literature, regardless of a large photon flux in the visible region, even near the core of MCs.[44] To date, even the absorption of visible light by $(H_2O)_n$-OH clusters has not been studied, neither experimentally nor theoretically, before our study; the interaction of ice dust with visible light should also have implications for atmospheric chemistry.

## 4.2 In-mantle chemistry of astronomical ice

Compared to the surface processes on astronomical ice dust, chemical processes within the ice mantle have not been thoroughly elucidated. In theoretical chemical models, the surface and bulk (i.e., in the ice mantle) processes are separated, and the latter is often assumed to be inactive.[45] There are models that include active bulk chemistry involving the diffusion of radicals.[46-48] In these models, only thermal diffusion processes are considered. Although proton transfer processes (i.e., the Grotthuss mechanism) are not active in low-temperature ice, the PHT mechanism (see Section 3) might contribute to the transportation or generation of $OH^-$ ions in the ice mantle. Considering the large proton affinity of the $OH^-$ ions (1633 kJ $mol^{-1}$), they will easily subtract protons from hydrogen-containing molecules once encountered; e.g., they will react with a hydrocarbon (RH) to produce a carbanion ($R^-$) according to $OH^-$ + RH → $H_2O$ + $R^-$.

We have recently investigated a possible in-mantle chemistry involving H atoms,[49] and important findings are briefly introduced here. Hydrogenation reactions such as reaction (1) are known to efficiently proceed on ice surfaces, but their efficiency significantly drops at temperatures above approximately 20 K. We found that, when reactants are buried deep in ASW, the hydrogenation reaction occurs even at 70 K and is attributed to a prolonged residence time of H atoms attained by trapping in cracks of ASW. Further verifications by theoretical chemical models are highly desirable for the evaluation of the impacts of in-mantle processes on chemical evolution in MCs.

■ **AUTHOR INFORMATION**

**Corresponding Author**

**Naoki Watanabe** − *Institute of Low Temperature Science, Hokkaido University, N19-W8, Kita-ku, Sapporo, Hokkaido 060-0819, Japan*; orcid.org/0000-0001-8408-2872; Email: watanabe@lowtem.hokudai.ac.jp

**Author**

**Masashi Tsuge** − *Institute of Low Temperature Science, Hokkaido University, N19-W8, Kita-ku, Sapporo, Hokkaido 060-0819, Japan*; orcid.org/0000-0001-9669-1288; Email: tsuge@lowtem.hokudai.ac.jp

**Author Contributions**

M.T. drafted the manuscript. Both M.T. and N.W. contributed to the final version of the manuscript.

**Notes**

The authors declare no competing financial interest.

**Biographies**

**Masashi Tsuge** received his Ph.D. degree from Tokyo Institute of Technology in 2007 for his spectroscopic studies on free radicals and weakly bound complexes. After joining the research group at ILTS in 2017, his research focuses on reaction dynamics relevant to the chemical reactions on/in interstellar grains.

**Naoki Watanabe** received a Ph.D. degree in atomic and molecular physics from Tokyo Metropolitan University in 1993. As a postdoctoral researcher at RIKEN until 1996, he worked on the photoionization processes of highly charged ions and molecules using synchrotron radiation. Since becoming a staff member at the Institute of Low Temperature Science, Hokkaido University in 1996, he has studied physics and chemistry on solid surfaces, especially solid water, at very low temperatures, which is relevant to chemical evolution in space. He is currently a professor at HU, and his research interest also covers gas phase reactions, including clusters at low temperatures.

## ■ ACKNOWLEDGMENTS

The authors thank the colleagues who contributed to the original works introduced in this paper and the technical staff at the Institute of Low Temperature Science for their contribution to the construction and renovation of the experimental apparatus. The works described here were supported by a JSPS Grant-in-Aid for Specially Promoted Research (JP17H06087) and partly by a JSPS KAKENHI grant (JP18K03717).